 \documentclass[final,1p,times]{elsarticle}

\usepackage[utf8]{inputenc} 
\usepackage{epsfig}    
\usepackage{amsmath}
\usepackage{palatino}
\usepackage{latexsym}
\usepackage{amssymb}
\usepackage{multirow}  
\usepackage{url}  

\usepackage{natbib,stfloats}
\usepackage{mathrsfs}
\usepackage{lscape}
\usepackage[usenames, dvipsnames]{color}
\usepackage{array}
\usepackage{graphicx}

\newcolumntype{P}[1]{>{\centering\arraybackslash}p{#1}}

\journal{\footnotesize and accepted in (POSTPRINT VERSION) IET Communications, Oct. 2020, vol. 14, Issue 17, pp. 3039-3046. ISSN: 1751-8628. \hspace{1cm}}

\begin{document}

\begin{frontmatter}

\title{Two-stage memetic algorithm for blind equalization in direct-sequence/code-division multiple-access Systems}

{\author[label1]{Luis M. San-Jos\'e-Revuelta\corref{cor1}}\ead{lsanjose@tel.uva.es}   \author[label1,label2]{Pablo Casaseca-de-la-Higuera}}

\cortext[cor1]{Corresponding author.  \hspace{1cm}  Final published version: DOI: 10.1049/iet-com.2019.0692.}

\address[label1]{Department of Signal Theory and Communications, University of Valladolid, Paseo de Bel\'en 15, 47011, Valladolid, Spain.}
\address[label2]{School of Computing, Engineering and Physical Sciences, University of the West of Scotland, High Street, Paisley, PA1 2BE, Scotland.}


\begin{abstract}
This paper proposes a novel memetic algorithm (MA)  for the blind equalization of digital multiuser channels with  Direct-Sequence / Code-Division Multiple-Access (DS/CDMA) sharing scheme. Equalization involves two different tasks, the estimation of: (1) channel response and (2) transmitted data. The corresponding channel model is first analyzed  and then the MA is developed for this specific communication system. Convergence, population diversity and near-far resistance  have been analyzed. Numerical experiments include comparative results with traditional multiuser detectors as well as with other nature-inspired approaches. Proposed receiver is proved to allow higher transmission rates over existing channels, while supporting stronger interferences as well as fading and time-variant effects. Required computation requisites are kept moderate in most cases. Proposed MA saves approximately 80\% of computation time with respect to a standard genetic algorithm and about 15\% with respect to a similar two-stage memetic algorithm, while keeping a statistically significant higher performance. Besides, complexity increases only by a factor of 5, when the number of active users doubles, instead of $32\times$ found for the optimum maximum likelihood algorithm.  The proposed method also exhibits high near-far resistance and achieves accurate channel response estimates, becoming  an interesting and viable alternative to so far proposed methods. 
\end{abstract}


\end{frontmatter}

\section{Introduction}
\label{sec_intro}

The use of wireless communications and applications has experienced a huge growth since the 90s. At the same time, transmission media (coaxial cable, optic fiber, etc) have also improved their technology and capacity. Users demand higher storage and transmission  capacity since both terminals and applications become more complex every year. Consequently, techniques such as Direct-Sequence /Code-Division Multiple-Access (DS/CDMA) that admit several users in a single RF channel bandwidth, have been widely investigated. This multi-code system enables telecommunication operators to offer high data rates  using available (limited) bandwidth \textcolor{black}{\cite{Proakis98,Verdu98,Hoshyar08,Cai18}}.

Its improvements in soft hand-offs, capacity, low probability of interception, interference rejection, low power spectral density, privacy and security makes of DS/CDMA an interesting scenario for testing novel algorithms \cite{Proakis98}. \textcolor{black}{CDMA is one of the most studied and implemented multi-access protocols. It was widely used in the second and third (2G and 3G) generations of cellular communications. In 5G, although it was initially one of the handled promising alternatives until a few months ago \cite{Cai18,Kumar20} –-specifically Low Density Signature (LDS) CDMA  \cite{Hoshyar08,Wang18,Zhang20}--, in the last available specification {\em 5G-NR Release 16}, 3GPP opted for OFDM-derived schemes \cite{Qualcomm20,3gpp20}. However, countless applications today use CDMA, which is still being considered in many areas of research. For instance, CDMA continues to be investigated and implemented in numerous Internet of Things (IoT) related applications  \cite{Hu16,Sagari17,Yassein17}, in submarine communications \cite{Kim19,Liu19,Petrosky19,Rahmati19}, in high-performance fiber optic communications \cite{Chang19,Ji20}, or satellite communications in low orbit systems (LEO) \cite{Papathanassiou01}. In modern systems where security is a key design aspect, DS/CDMA has been proposed combined with both  chaotic spreading sequences \cite{Rahnama13,Kaddoum16,Divya18,Yao18,Liu19,Tutueva20,Kumar20b} and with MIMO techniques \cite{Mehrizi17} to decrease the level of MAI, improving, this way, previous TDMA, SDMA and conventional CDMA schemes.}


Modern systems require very high transmission rates. At these rates, multi-access interference (MAI) and near-far effects are the two main critical issues affecting DS/CDMA cellular networks \cite{Proakis98,Lain07}. Intersymbol and multi-access interferences (ISI and MAI) must be taken into account and controlled so that performance does not degrade \cite{Proakis98}. Both ISI and MAI can be controlled from many points of view. Conventional multiuser detectors (MUDs) make use of filters matched to the codeword of the user of interest. However, this is only optimum if all received codewords are independent. In real applications this situation rarely occurs and performance notably reduces, especially if near-far degradations are present. In 1986, S. Verd\'u studied this issue and found that the joint extraction of users transmitted sequences could mitigate this problem \cite{Verdu98}. However, complexity of the optimum detector relying on the ML criterion grows exponentially with the number of transmitters, turning unfeasible in real scenarios  \cite{Proakis98,Verdu98}. Therefore, many authors have paid attention to the development of suboptimal schemes that can be used on real systems. Several techniques based on Natural Computation and Artificial Intelligence have been proposed to address multiuser detection (see next Section).

This paper deals with the development of a novel two-stage memetic algorithm (MA) and its application to blind joint channel estimation and symbol detection in DS/CDMA systems. \textcolor{black}{Explicit estimation of channel coefficients allows posterior processing tasks such as MIMO signal separation, compensation of distortions and computation of certain channel parameters that can be fed back to the receiver for transmission fine-tuning.} Proposed MA efficiently solves the CDMA equalization problem, while overcoming  the drawbacks found in other approaches.  For instance, conventional evolutionary algorithms (EAs) can only estimate the optimum search space area within a cost-effective time and present great problems in fine-tuning solutions \textcolor{black}{\cite{Khan15,SanJose16,Huang17,Asif19,Iqbal19}}. These disadvantages can be overwhelmed by  applying exploitative search to optimize the final population of solutions estimated by the EAs \textcolor{black}{\cite{Lim04,Cotta18}}. Both effectiveness and solutions' quality are enhanced using this two-step approach. This  big family of optimization methods, which have elements from metaheuristic and evolutionary algorithms, and also include local learning or improvement procedures are generally known as {\em memetic algorithms} (MAs) \cite{Cotta18}. These have a number of advantages, such as simple implementation  and capability to deal with different functional problem representations \textcolor{black}{\cite{Lim04,Cotta18}}.

In our case, a two-stage algorithm is used. First, a genetic algorithm with a specific method for adjusting diversity is used to find the optimal search area within a tolerable time. Secondly, a procedure making use of the $k$-opt heuristic local search algorithm is used to improve the solution estimated during first stage, and thus extract the global optimum solution with low computational expense \cite{Kernighan70,Lin73}. The main novelties of our proposal are: (i) the first stage uses both mutation and crossover (the latter with a very low probability), as opposed to \cite{Lim04}, where mutation was only used, (ii) population diversity is monitored and controlled using the fitness entropy, (iii) probabilities of mutation and cross over are on-line adjusted using the fitness entropy, (iv) an elitism strategy is introduced in the first stage. Besides, mutation is applied to part of this elite, (v) a simple to implement termination criterion, \textcolor{black}{and (vi) high bandwidth efficiency since it does not requiere training sequences}. These improvements allow to work with smaller population sizes and less iterations, while convergence and near-far resistance are improved.

The remainder of this paper is structured as follows: literature review is shown in section \ref{litrev}, whilst section \ref{sec_mud} explains equalization --joint symbol and channel estimation-- in DS/CDMA systems. Basic concepts and notation are here presented, along with the proposed fitness function to be used in the MA. Section \ref{sec_dm} describes the proposed MA and shows how population diversity is monitored and controlled using the population fitness entropy. Next, section \ref{sec_nr} shows the numerical results with emphasis on the comparison with other multiuser detectors, both traditional and nature-inspired.  Finally, conclusions are presented in section \ref{sec_concl}.

\section{Literature review}
\label{litrev}

Many different techniques based on Natural Computation and Artificial Intelligence have been proposed to address multiuser detection. Initial approaches were based on single-step evolutionary algorithms (EAs), mainly Genetic Algorithms (GAs). The work in \cite{Juntti97} proposed a synchronous DS/CDMA MUD based on a GA. It is based on AWGN channel and does not use diversity techniques. Its main drawback is the requirement of good estimates of the first transmitted symbols. Later, in 2004, Yen studied the  asynchronous case in \cite{Yen04}, where the effect of the surrounding symbols from other system users is taken into account. This algorithm also estimates those symbols that are adjacent to those from the user of interest. Another variant of the GA, whose performance is close to the optimal by introducing a local search algorithm before the GA,  was proposed in \cite{Yen00}. This idea of adding modules to the standard GA was used also in \cite{Ergun00}. In that case a multistage detector is integrated into the GA in order to speed  convergence up.

Some remarkable recent works using GAs have tried to estimate the channel response, most of them are focused on selective Rayleigh channels \cite{Tan10,Maradia09}. On the other hand, some approaches have studied both bit-error-rate (BER)  and near-far effect performances. It is worth mentioning recent approaches using GAs such as \textcolor{black}{\cite{Nooka13,Rashid13,Huang17,Khan15,SanJose16,Iqbal19}}. Earlier references can be found in  \cite{SanJose05}.

Apart from GAs, several other nature-inspired methods have also been applied to multiuser detection.
During last decade it deserves special attention those bio-inspired methods based on swarm and evolutionary computation. These methods include the use of ant-colony algorithms (ACO) \cite{Hijazi04,Lain07,Marinello12,Xu07} and particle swarm optimization (PSO) \textcolor{black}{ \cite{Arani13,Oliveira06,Soo07,Wang14,Kaur16,Asif19} }, among others. \cite{Arani13} combines a PSO algorithm with the conventional detector, which is initially used to initialize the position of a particle, showing, this way, better capability against bit error tan conventional detector. In \cite{Lu04} a binary PSO is applied for CDMA multiuser detection, while \cite{Soo07} proposed the use of the decorrelating detector and the linear minimum mean square error (MMSE) detector to initialize the PSO multiuser detector. Recently, \cite{Wang14} proposed a completely binary PSO algorithm which is reported to resist higher noise levels and to converge in a very short time. \textcolor{black}{\cite{Kaur16} compares performance of a PSO-based detector for CDMA with another receiver that uses the Spider monkey optimization (SMO) scheme, showing better results for the later}. \cite{Larbi14} includes an interesting list of references in this area.

This paper proposes the use of a memetic algorithm to solve the equalization problem in DS/CDMA, i.e. the joint symbol detection and channel estimation tasks in a multiuser communication system.  To the best of our knowledge, no work has been dedicated to the joint estimation of fading coefficients and users' data symbols using a MA.

\section{Channel estimation and symbol detection in DS/CDMA}
\label{sec_mud}

\subsection{DS/CDMA channel model}
\label{dscdma}

The multiuser communication environment and the channel model used in this work is next described. This channel is simultaneously shared by $U$ active users transmitting binary symbol sequences. Each user operates with a confidential normalized signature (or {\em codeword}) from set $\{s_i(t)\}_{i=1}^U$. Channel response is considered to be completely represented by both a set of flat-fading coefficients, and an additive zero-mean white Gaussian noise (AWGN) component. All signals are supposed to be synchronously transmitted. This assumption is considered for simplicity as it captures most of the effects of asynchronous systems with a low delay spread \cite{Lamarre12}.

User $i$ transmits an $F$-length sequence $x_i(n)$ of statistically-independent symbols. Each of them  modulates a codeword, $s_i(t)$, which is obtained as
\begin{eqnarray}
    s_i(t)=\sum_{\ell=0}^{N-1} s_{i,\ell} \gamma (t-\ell T_c)
\label{sit}
\end{eqnarray}
where ${\bf s}_i=(s_{i,0},...,s_{i,N-1})^T$ stands for the signature of user  $i$, $T_c=T/N$ is the chip period ($N$ is known as {\it processing gain}),  $T$ denotes the symbol period  and $\gamma(t)$ represents a chip waveform whose energy has been normalized.

As a consequence, the frequency content is spread by a factor $N$ and the original narrowband signal is de-sensitized to some potential channel degradation and interference \cite{Proakis98}. This way, the $i$th user transmits the following signal
\begin{eqnarray}
    y_i(t) = \sum_{n=0}^{F-1} x_i(n) s_i(t-nT)
\label{xit}
\end{eqnarray}
where $x_i(n)$ are the data symbols transmitted by the $i$th user. The signal at the receiver input is
\begin{eqnarray}
    r(t) &= \sum_{i=1}^U r_i(t) + g(t)     \qquad    0 \leq t \leq T_F
\label{rt}
\end{eqnarray}
where $g(t)$ denotes a complex AWGN component, which is not correlated with the users' transmitted symbols $x_i(n)$, $T_F$ represents the frame duration, and $r_i(t)$ is
\begin{eqnarray}
    r_i(t) = \sqrt{E_i} \sum_{n=0}^{F-1}  b_i(n) x_i(n) s_i(t-nT)
\label{rt2}
\end{eqnarray}
$E_i$ denotes the energy per bit of the $i$th user, and $b_i(n)$ the flat-fading coefficient of this user. In this work, a non-stationary channel is considered, with a time variation  model defined in \cite{Yen01}, where fading coefficients $b_i(n)$ change with time as a function of a Doppler frequency, $f_d$, as
\begin{eqnarray}
    b_i(n+1) = \alpha \cdot b_i(n) + \varphi
\end{eqnarray}
with $\alpha=\exp(-2\pi f_d T)$ and $\varphi$ represents a zero-mean AWGN signal.

The joint task of channel response estimation and detection of transmitted symbols can be seen in Eq. (\ref{rt2}), where both data symbols $x_i(n)$ and fading coefficients $b_i(n)$ need to be estimated.

\begin{figure*}[htb]
\centerline{\psfig{file=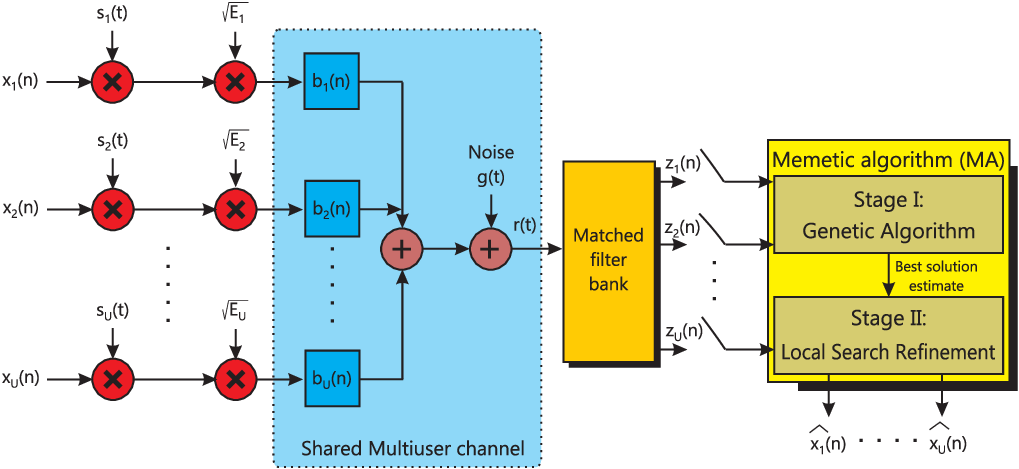,width=0.85\textwidth}}
\caption{Representation of the multiuser channel model where $U$ users transmit simultaneously using a DS/CDMA sharing strategy. $E_i$: bit energy of user $i$, $x_i(n)$: symbol sequence transmitted by user $i$, $s_i(t)$: codeword of user $i$, $\hat{x}_i(n)$: estimate of $n$th symbol transmitted by user $i$.}
\label{fig_channel}
\end{figure*}

An schematic representation of this multiuser communication system is shown in Fig. \ref{fig_channel}. The first block in the proposed receiver consists of a set of $N$ filters distributed in parallel, each of them matched to a different user signature, just after sampling the received signal at $1/T$ rate. The aim of this bank of filters is to capture signal energy only from the user of interest.


In order to estimate the transmitted symbols' vector ${\bf x}(n)=[x_1(n),\dots,x_U(n)]^T$, we have followed ideas in \cite{Yen01}, where estimation is presented as a maximization issue, and the output of the matched filters' bank, ${\bf z}(n)$, is obtained as
\begin{eqnarray}
    {\bf z}(n)=[z_1(n),\dots,z_U(n)]^T = {\bf R} {\bf B}(n) {\bf E} {\bf x}(n) + {\bf g}
\end{eqnarray}
with ${\bf R}$ being the  cross-correlation matrix of users' codewords, ${\bf B}(n)=$ $\mbox{diag}(b_1(n),\dots,b_U(n))$, ${\bf E}=$ $\mbox{diag}(\sqrt{E_1},\dots,\sqrt{E_U})$, ${\bf x}(n)=$ $[x_1(n),\dots,x_U(n)]^T$ and ${\bf g}=$ $[g_1(n),\dots,g_U(n)]^T$.

In \cite{Fawer95} it is demonstrated that, given vector ${\bf z}$, the log-likelihood conditional pdf given both the fading coefficients' matrix and the transmitted symbols' vector, can be obtained as
\begin{align}
    {\mathcal L}\left( {\bf B}(n), {\bf x}(n)  \right)  =  &2 \Re \left\{ {\bf x}(n)^T {\bf E} [{\bf B}(n)]^* {\bf z}(n)  \right\}     \nonumber   \\
        &-  {\bf x}(n)^T {\bf E} {\bf B}(n) {\bf R} [{\bf B}(n)]^* {\bf E} {\bf x}(n)
\label{metrica}
\end{align}
with ``$\Re$'' and ``*'' stand for the real part of a complex magnitude, and the complex conjugate operator, respectively. Hence, both the fading coefficients' matrix and the users' transmitted symbols are estimated as
\begin{eqnarray}
    (  \widehat{{\bf B}(n)},  \widehat{{\bf x}(n)}   )  =    \arg \max_{{\bf B}(n),{\bf x}(n)}  \left\{ {\mathcal L} \left( {\bf B}(n),{\bf x}(n) \right)  \right\}
\label{maxprob}
\end{eqnarray}

Finally notice that fading coefficients are supposed to vary slowly enough so that  this fading is assumed to be constant within each symbol interval. Besides, fadings from users $i$ and $j$, $i \neq j$, are considered to be independent.

As a consequence, the optimization problem to solve consists in the joint estimation of ${\bf B}(n)$ and ${\bf x}(n)$, which is repeated at every symbol period $T$.

\section{Design methodology}
\label{sec_dm}

The proposed algorithm involves two stages: first, it uses a GA with a diversity control procedure, and afterwards,  an heuristic $k$-opt exploitative search algorithm to fine-tune the output from the first stage. It is well known that one of the main advantages of GAs is that, considering a specific application, they require very few a priori assumptions to achieve a near optimum estimate \cite{Ergun00,Juntti97,Yen04}.  This GA constitutes the first stage. Next, the local search method fine-tunes the fittest GA output so as to efficiently provide a quasi-optimum solution. Notice that this procedure is run assuming that the first stage (GA) has reached a near-optimum solution estimate.

\subsection{Genetic algorithm (Step 1)}
\label{sec_ga}

\subsubsection{Basic concepts}
\label{sec_ga_fund}

In a GA, possible solution estimates are encoded in a binary vector usually known as {\em chromosome} (or {\em individual}) and, during the GA cycle, {\em genetic operators} are applied to a subgroup of the fittest chromosomes with the aim of preserving crucial knowledge. The selection process relies on the principle of {\em survival of the fittest individuals}: those ones with highest fitness will have more chances to be selected for reproduction.

This description of the standard GA is quite generic, however, implementation of many aspects can be particularized to each specific problem, for instance: initial generation of individuals and their encoding, definition of the operators (selection, crossover and mutation) and many other implementation tasks. Some of these design issues are next described.

\subsubsection{Coding, selection and fitness evaluation}
\label{sec_fitness}

First, the initial population is randomly created with $n_p$ binary-encoded individuals, and each one is evaluated according to its fitness. In the proposed equalization problem, Eq. (\ref{metrica}) is used for evaluating the fitness of every member of the population, since both transmitted symbols and fading coefficients are encoded into chromosome $\emph{\bf CHR}_i(n,k)$ as:
\begin{align}
     \emph{\bf CHR}_i(n,k) &= [ {\bf B}_i(n,k) ,  {\bf x}_i(n,k)  ]  \nonumber \\
   &=    [  (b_{i,1}(n,k),b_{i,2}(n,k), \dots , b_{i,U}(n,k) )  , \dots \nonumber \\  & \dots (x_{i,1}(n,k),x_{i,2}(n,k), \dots , x_{i,U}(n,k))  ]    \nonumber \\
   &  1 \leq k \leq n_g, \hspace{0.2cm}  1 \leq i \leq n_p, \hspace{0.2cm}  0 \leq n \leq F-1
\label{ui}
\end{align}
where  $n$, $k$, $i$ and $U$ stand for the $n$th symbol period, the $k$th GA generation, the $i$th population individual,  and the amount of active transmitters, respectively. On the right part of $\emph{\bf CHR}_i(n,k)$, vector ${\bf x}_i(n,k)$ represents the estimates of the $U$ transmitted symbols. On the left, fading coefficients $b_i(n,k)$ are encoded. Since these coefficients are usually complex values, their real and imaginary parts are, each one, binary encoded, using 10+1 (sign) bits. A 22-bit long string is this way obtained. 

Due to the assumption of synchronous transmission, one GA is executed in each symbol interval (every $T$ seconds). Therefore, fading coefficients estimates obtained when the GA finishes, are kept as initialization values for next symbol period, where a new GA will start. This agrees with the hypothesis of low time variation of the channel fading, involving that fading coefficients slightly change within each period $T$.

In contrast to the MA proposed in \cite{Lim04}, our proposed genetic algorithm implements two different genetic operators: crossover and mutation, with the former used with a very low probability.

Selection of chromosomes for mutation and crossover is based on a stochastic rule, where the fittest chromosomes can be selected with a higher probability than the weakest ones. Therefore, the $i$th chromosome $\emph{\bf CHR}_i(k)$ will be selected  at iteration $k$ with probability \textcolor{black}{$\Phi_i(k) / \sum_{j=1}^{n_p} \Phi_j(k)$}, where  $\Phi_i$ represents the fitness of individual $i$ given by Eq. (\ref{metrica}). This scheme is readily implemented using a selection method known as {\it roulette wheel}, where the size of each circular sector is proportional to the aptitude of each chromosome \cite{Mitchell98}.

\subsubsection{Genetic operators}
\label{sec_genop}

Two genetic operators are applied: mutation and crossover. Mutation modifies chromosomes with probability $P_m$, modifying the value of certain positions. Both the specific position within $\emph{\bf CHR}_i$ and its new value, are randomly obtained. This operator increases the explorative search of the solutions' space. A low probability $P_m$ prevents any part in $\emph{\bf CHR}_i$ from remaining fixed, while a high value results in a random search. Therefore, $P_m$ should be correctly adjusted. For instance, in our application, and in accordance to \cite{Lai96}, a nice trade-off solution is achieved with an initial value $P_m(0) \in [0.02-0.05]$.

Mutation operator is invoked following the scheme proposed in \cite{Lim04}, i.e., descendant chromosome comes given by
\begin{equation}
    \emph{\bf NEW\_CHR}_{i,k}=sign(\emph{\bf CHR}_{i,k} + \mathcal{N}_k(0,\sigma)),\quad k=1,2, \dots ,K
\end{equation}
where $\emph{\bf NEW\_CHR}_{i,k}$ and $\emph{\bf CHR}_{i,k}$ represent the $k$-th component of the $i$-th offspring and parent individuals, respectively, $\mathcal{N}_k(\eta,\sigma)$ is a Gaussian random variable with mean $\eta$ and standard deviation $\sigma$. A new  random value is obtained for each $k$. Notice that $\sigma$ allows to control how closely the offspring is relative to its corresponding parents.

Meanwhile, crossover operator obtains two new chromosomes ({\it descendants}) by merging two parents at specific points. The crossover operator is applied with probability $P_c \ll 1$. Since progenitors are selected from highest fitness chromosomes, the small modifications performed over these data structures are supposed to generate good individuals, as well.

Crossover is applied with probability $P_c = 0.01$. We get next generation by selecting the $n_p$ fittest individuals from the parents and descendants sets considered together.

\subsubsection{Elitism, termination criteria and convergence}
\label{sec_elitism}

Elitism is implemented in our GA. This means that individuals with the highest fitness are directly selected and inserted into next generation. In our experiments mutation is applied to half of the elite with probability $P_{m,e}=0.2 P_m$, while crossover is not implemented.

Successive iterations of the GA are performed until a termination criterion is verified; in our real-time DS/CDMA application, the algorithm is iterated until a predetermined number of fitness evaluations is reached. Once the last iteration has been run, the GA output is given by the chromosome whose fitness is the highest,
\begin{multline}
    \emph{\bf GA\_OUTPUT} =  \emph{\bf CHR}_{i\_best} (n_g,k)  = \\
    = [ {\bf B}_{i\_best}(n_g,k) ,  {\bf x}_{i\_best}(n_g,k) ]
\label{sol_ga}
\end{multline}

\begin{figure}[!htb]
\centerline{\psfig{file=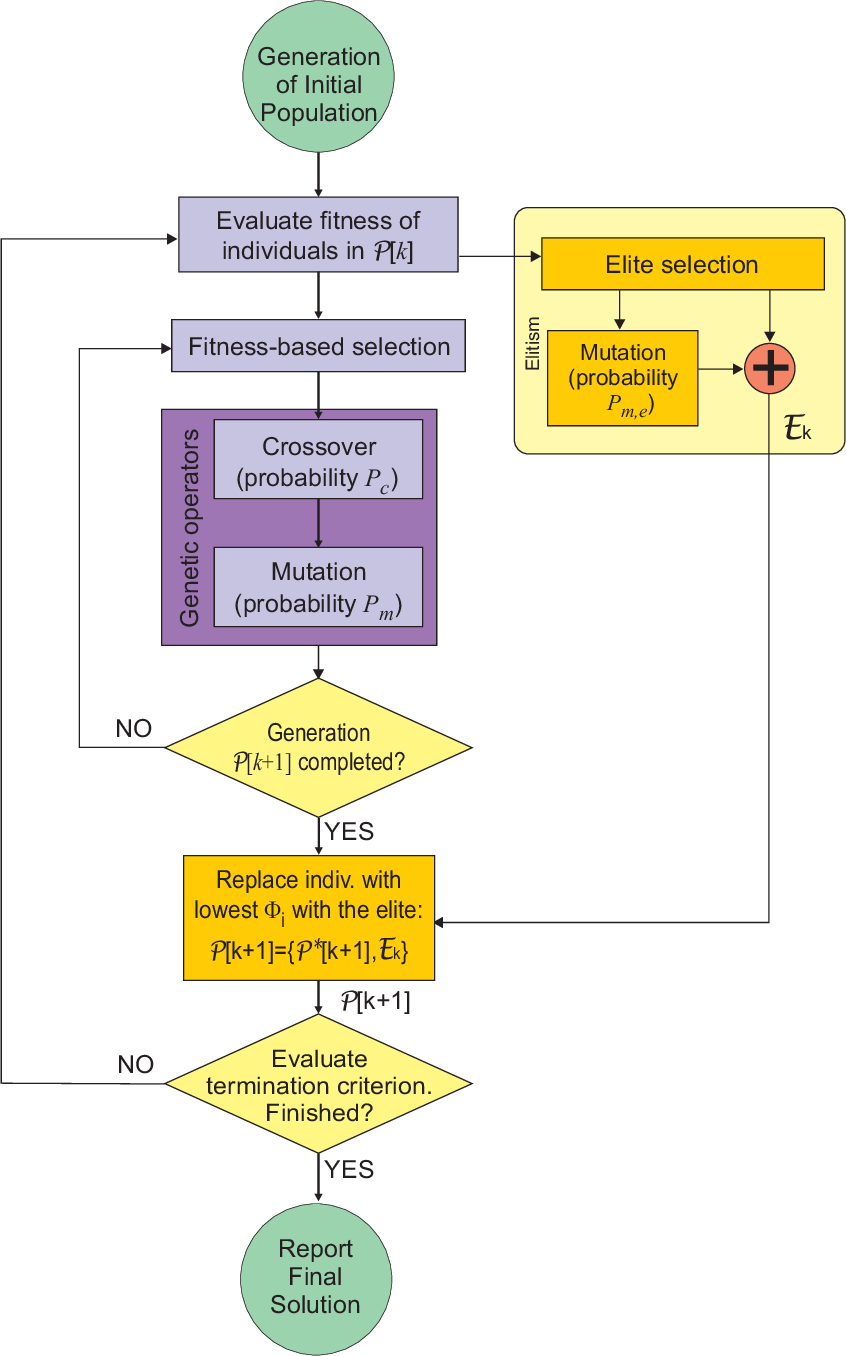,width=7cm}}
\caption{Simplified flowchart of the proposed GA constituting the first stage of the memetic algorithm. Two genetic operators are implemented: crossover and mutation. Elitism is also implemented at the end of each iteration.}
\label{GAflow}
\end{figure}

An schematic flowchart of the GA structure is shown in Fig. \ref{GAflow}.

Once the GA has finished, the exploitative searching method of Step-2 is invoked to improve the solution estimate in Eq. (\ref{sol_ga}).

\subsubsection{Diversity control}
\label{sec_div}

GA convergence is greatly improved with the introduction of procedures that optimize population diversity. Classic GAs tend to converge to non-optimal solutions, mainly as a result of a selection that greatly relies on fitness \cite{Ursem02}. This implies that population will be formed mainly by the fittest individuals, resulting in low diversity and low quality solutions.
On the other hand, GAs suffer from excessive computational load, in part due to the load of the genetic operators in addition to fitness evaluations. Common population sizes ($n_p$) can be a huge number of several hundreds (300-2000) for equalization involving both users' data and channel coefficients estimation. The GA here proposed works with less chromosomes (60 to 400 individuals) due to the use of a population diversity control scheme, where genetic operators rely on the Shannon entropy of the population fitness, which is obtained as
\begin{eqnarray}
    {\mathcal H}({\mathcal P}[k]) = - \sum_{i=1}^{n_p} \Phi_i^* \log \Phi_i^*
\label{entropia}
\end{eqnarray}
with  $\Phi_i^*(k)$ being the normalized value of the fitness of the $i$th individual, i.e.,
\textcolor{black}{
\begin{eqnarray}
     \Phi_i^*(k) = \frac{\Phi_i(k)}{\sum_{j=1}^{n_p} \Phi_j(k)}, \qquad 1 \leq i \leq n_p
\label{entropianorn}
\end{eqnarray}
}
The aim is to adapt the explorative/exploitative sense of the search depending on the population diversity estimated at each stage of the convergence cycle. 
When entropy ${\mathcal H}$ is high, it means that population individuals are very similar. In this case, $P_m$ is increased and $P_c$ is decreased, in order to increase diversity by boosting explorative search. On the other hand, if ${\mathcal H}$ is low, meaning that individuals are quite different, then $P_m$ is decreased and $P_c$ increased.

The thus designed GA is implemented in stage I of the proposed MA. It requires notably less computational load than the standard GA since dynamic operators allow to work with  smaller population sizes, and, consequently, genetic operators and fitness evaluations are computed less frequently.

\subsection{Local refinement using the $k$-opt algorithm (Step-2)}

Local search procedures focus their search in the neighborhood of the best solution estimate found so far  until no improvement is found. The {\em neighborhood} of chromosome $\emph{\bf CHR}_i$ is defined as the set of chromosomes obtained by flipping one or more components in the chromosome encoding --see Eq. (\ref{ui}).

The smallest neighborhood is obtained flipping only one bit of the chromosome. This is known as {\em 1-opt neighborhood}, and aims to find the position whose flip involves the highest fitness gain $(\emph{GAIN}=\Phi({\emph{\bf CHR}_i}')-\Phi({\emph{\bf CHR}_i}))$ in each iteration. Lim et al. estimated the gain associated to the change of value of the $k$th bit in a chromosome \cite{Lim04}.

In general, the $k$-opt neighborhood of a specific chromosome is the set of chromosomes obtained by simultaneously flipping $j$ bits ($1\leq j \leq k$) in that chromosome. This can be expressed as:
\begin{equation}
    \emph{NH}_{k-opt}(\emph{\bf CHR}_i) = \left\{ \emph{\bf CHR}_i' \in \mathcal{S} / d^*(\emph{\bf CHR}_i,\emph{\bf CHR}_i') \leq k \right\}
\end{equation}
where $\mathcal{S}$ represents the complete space of solutions and $d^*$ stands for the Hamming distance between two binary vectors. Notice that, the $k$-opt neighborhood size grows exponentially with the number of flipped bits $k$: $|\emph{NH}_{k-opt}(\emph{\bf CHR}_i)| = \sum_{i=0}^k {k \choose i}$.

The principles of the Lin-Kernighan algorithms can be used to  find a subset of the $k$-opt neighborhood \cite{Kernighan70,Lin73}. The goal is to improve  solution by changing certain positions of it in every iteration. A set of $K$ new estimates ($K$ is the length of the binary vector $\emph{\bf CHR}$) is obtained  by changing the value of the bit whose associated gain is the highest. Each bit can only be changed once. 
This yields a set of $K$ solutions, from which only the best one becomes the input for next iteration. This way, the value of a variable number of positions is changed in each iteration in order to improve the solution quality.

\section{Results and Discussion}
\label{sec_nr}

Performance of the developed algorithm is next evaluated using numerical simulations. Synchronous transmission and additive white Gaussian noise is considered unless otherwise specified. Binary Phase Shift Keying (BPSK) modulation, a squared chip waveform  $\gamma(t)$ --see Eq. (\ref{sit})--, a processing gain $N=31$ and a Doppler frequency $f_d = 1$ krad/sec, are used. Besides, orthogonal signatures are conformed using Gold sequences. All algorithms were tested on an Intel(R) Core(TM) i5-6200U CPU (2.4GHz) and 8G RAM with Matlab(R).

\subsection{Probability of Error vs SNR}

\begin{figure}[htb]
\centerline{\psfig{file=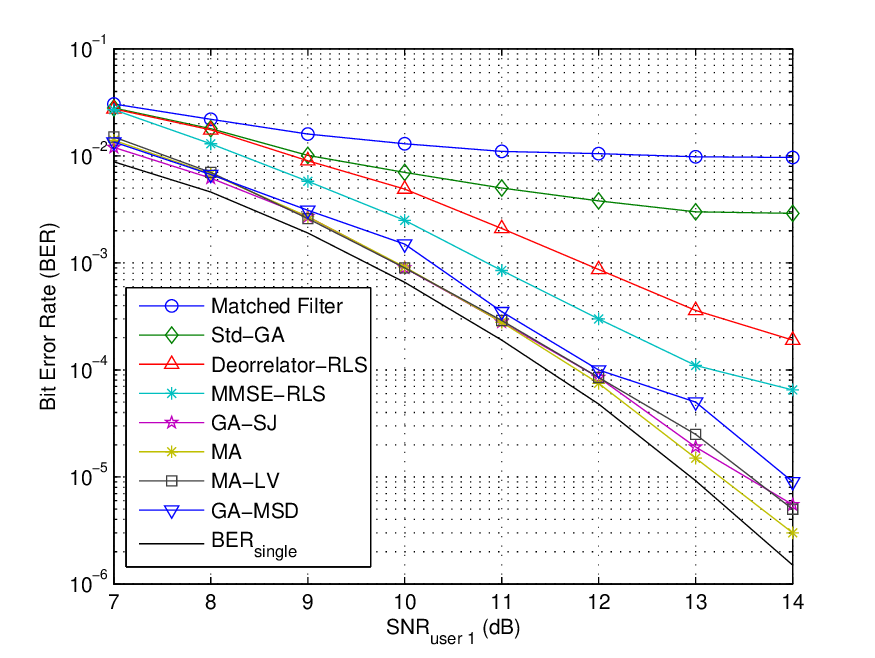,width=9.5cm}}
\caption{Estimation of the Bit Error Rate performance for different values of the signal-to-noise ratio (SNR, dB) for several multiuser detectors. $U=10$ active users. Near-far distortion: $E_j/E_1$=4 dB for $j > 1$.} 
\label{fig_ber_snr1}
\end{figure}

Fig. \ref{fig_ber_snr1} shows the bit error probability (BER) of User 1 (also known as {\em user of interest}, UOI) as a function of  signal quality, which is evaluated using the signal-to-noise (SNR) power ratio $E_1/N_0$, for different estimation algorithms. $U=10$ active users and a near-far effect of 4 dB (i.e. $E_j/E_1$ = 4 dB, $2 \leq  j \leq U$) representing that the remaining users have a 4 dB higher power, is considered. The first stage of the MA is implemented with a population size of $n_p$=60 chromosomes. 

The algorithms chosen for comparison are: (i) standard GA --though it is a simpler one-stage algorithm it is a commonly used key reference to compare with--, (ii) a recently developed GA (GA-SJ from now on) \cite{SanJose16}, (iii) the two-stage memetic algorithm developed by Lim and Venkatesk (MA-LV from now on) \cite{Lim04}, (iv) the two-stage detector by Ergun and Hacioglu \cite{Ergun00} (GA-MSD), (v) several conventional detectors, such as the matched filter (MF) detector, the decorrelator detector and the MMSE detector.

Comparing MA to a standard GA-based detector, it can be seen how MA's performance approaches that of the single user scenario (a lower bound for the BER of any multiuser detector is obtained by estimating the probability of error when $U=1$ --i.e. absence of interfering users-- \cite{Proakis98}).  GA-based MUD has difficulties to converge as SNR increases, even when implemented with a higher number of iterations and generations. Specifically, the proposed MA is implemented with $n_{p,MA}=60$ individuals, while GA uses $n_{p,GA}=300$. On the other hand, the number of iterations in GA is $n_{g,GA}=500$, while step-1 of our MA uses 250.

The GA-SJ algorithm is reported to be an efficient method  \cite{SanJose16}. We can see how the BER achieved with the here proposed MA is very close to GA-SJ's BER. However, the introduction of a second stage in our MA allows to reduce the population size required in Stage 1 in about 20-25\%. As a consequence, global computation time is now reduced about $30\%$ with respect to GA-SJ.

This reduction in generations and population size involves a decrease in the number of required fitness function evaluations  (symbol $\sharp$ stands for {\em Number of}):
\begin{eqnarray}
        \sharp Fit\_Eval_{GA} &=& (1 + P_c + P_m) n_{g,GA} n_{p,GA}   \nonumber  \\
        \sharp Fit\_Eval_{MA} &\simeq& (1 + P_c + P_m) \frac{1}{3} n_{g,GA} \frac{1}{5} n_{p,GA} \nonumber \\
                              &=&  \frac{1}{15}  \sharp Fit\_Eval_{GA}  \nonumber \\
        \sharp Fit\_Eval_{GA-SJ} &\simeq& (1 + P_c + P_m) \frac{1}{4} n_{g,GA} \frac{2}{5} n_{p,GA} \nonumber \\
                              &=&  \frac{1}{10}  \sharp Fit\_Eval_{GA}   \nonumber \\
        \sharp Fit\_Eval_{MA} &\simeq& \frac{2}{3} \sharp Fit\_Eval_{GA-SJ}   \nonumber
\end{eqnarray}
However, due to the load of step-2 in the proposed MA, the final reduction in time is approximately 77\% with respect to Std-GA, and about 20\% with respect to GA-SJ.

Apart from one-stage GAs, two different nature-inspired methods were implemented for comparison, MA-LV \cite{Lim04} and GA-MSD \cite{Ergun00}. Fig. \ref{fig_ber_snr1} also shows results when both algorithms are implemented keeping an equivalent computational load (by properly setting the number of fitness evaluations). It can be seen that performances of MA and MA-LV are similar for ${\mbox{SNR}}_{\mbox{{\small user 1}}} \leq 11$ dB. For higher values proposed MA performs better. This is due to the fact that MA efficiently combines the explorative/exploitative sense of search due to the dynamic functioning of the genetic operators, by monitoring the population fitness entropy, as explained in section \ref{sec_div}. On the other hand, GA-MSD shows slightly higher BERs, specially for ${\mbox{SNR}}_{\emph user 1} > 12$ dB. Notice that performances of these three two-step algorithms are always better than those of conventional and one-stage methods. BER plots of three traditional detectors --the matched filter (MF) detector, the Minimum Mean Square Error (MMSE) detector and the decorrelator detector--, are shown in Fig. \ref{fig_ber_snr1}, as well.

Numerical results indicate that simple methods (standard GA and MF) cannot correctly converge even when the power of user 1 is high. These results are coherent since MF is known to suffer from BER performance degradation in fading channels \cite{Proakis98}. On the other hand, 
  MMSE and decorrelator lead to an intermediate performance between the two-stage methods and the MF or the standard GA.

Finally, we compare performances of three two-stage nature-inspired methods using the Friedman test \cite{Friedman40}. The problem test suite consists of 8 optimization problems, each one defined with a specific value of $\emph{SNR}_{\emph user 1}(dB) = \gamma$ dB, $\gamma \in [7,14]$. The Friedman test is conducted to detect significant behaviour differences between two or more detectors. Average rankings are shown in Table \ref{tabla_X}. The best average ranking (lowest value) is in italic and corresponds to our MA approach, which outperforms the other two schemes. A $p$-value of 0.0015 is obtained showing that there exist significant differences between the behaviour of the three algorithms.

\begin{table}[!htb]
\centering
\caption{Average ranking achieved with the Friedman test for three two-stage nature-inspired multiuser detectors. Each test problem comes given by a specific $\emph{SNR}_{\emph user 1}(dB)$ value, from set $\{7,8,9,10,11,12,13,14\}$.}
\begin{tabular}{c c}
   Algorithm     &    Friedman Test Score   \\
\hline
Memetic Algorithm  &   {\em 1.25}   \\
MA-LV              &   1.75  \\
Std-GA             &   3.00  \\
\hline
\end{tabular}
\label{tabla_X}
\end{table}

\subsection{Bit Error Rate {\it  vs} number of transmitting users (capacity)}

\begin{figure}[htb]
\centerline{\psfig{file=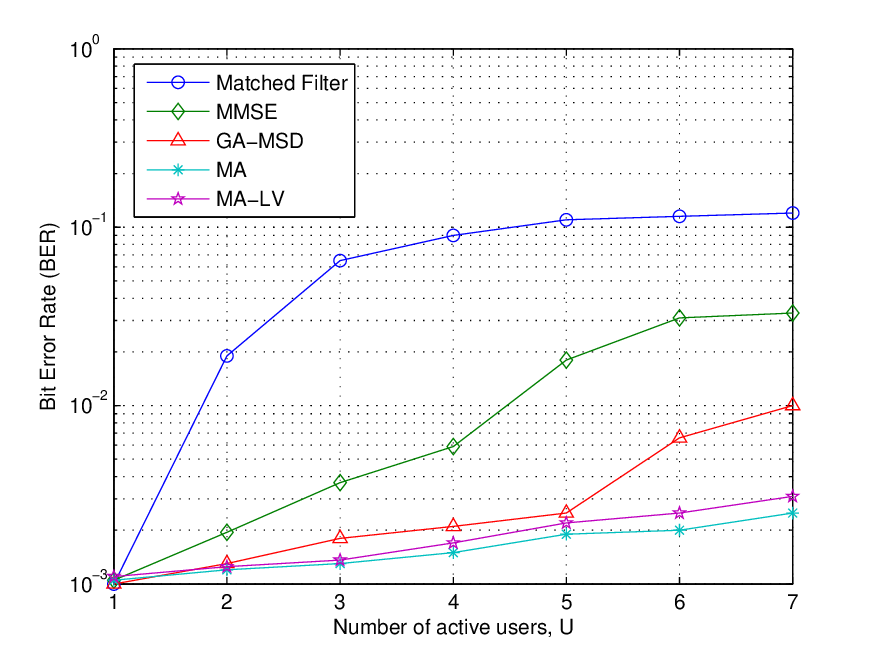,width=9.5cm}}
\caption{Estimates of the Bit Error Rate depending on the number of transitting users $U$. $E_i = E_j, 1 \leq  i,j  \leq U, i \neq  j$.} 
\label{fig_ber_k}
\end{figure}

The number of actively  transmitting users over the channel affects performance. The {\it capacity} of the channel quantitatively evaluates its ability to deal with this. Capacity curves show the bit error rate in terms of the number of active users. Fig. \ref{fig_ber_k} shows the plots for different methods (MA, MA-LV, GA-MSD, MF and MMSE), depending on the number of simultaneously transmitting users.

In this simulation every user is forced to transmit with the same power, i.e. $E_i = E_j, 1 \leq  i,j  \leq U, i \neq  j$. Simulation results show that nature-inspired schemes perform better than MF and MMSE, whose performances are clearly degraded by multi-access interferences.

Performances of MA and MA-LV (two-stage nonlinear schemes) are very close. However, MA's performance is a little better, specially when the number of users is high ($U>5$) and multi-access interference is more relevant. On the other hand, the GA-MSD detector shows a performance that drops with respect to MA and MA-LV. This degradation increases with the number of active users. This way, GA-MSD has an intermediate performance between two-stage MAs and conventional detectors.

Another notable advantage of the proposed MA and the other nature-inspired approaches,  in contrast to traditional optimization methods, is that they do not need any memory elements, since information related to previous iterations does not need to be stored.


Computational load in terms of the number of transmitting users $U$ is shown in Table \ref{tabla2}. Results concerning the standard GA (Std-GA), GA-SJ \cite{SanJose16} and MA-LV \cite{Lim04} multiuser detectors are included for comparison, as well.  Both Population size and the number of iterations are shown for $U=10$, 15 and 20.

\begin{table}[!htb]
\centering
\caption{Population size ($n_p$) and number of generations ($n_g$) required for different values of parameter $U$  (active users) for the following multiuser detectors: (i) proposed memetic algorithm (MA), (ii) standard genetic algorithm (Std-GA), (iii) GA-SJ \cite{SanJose16} and (iv) MA-LV.}
\begin{tabular}{ccccccccc}
\hline
      No. of       &    \multicolumn{4}{c}{Population size}   &    \multicolumn{4}{c}{Generations} \\
      users        &    \multicolumn{4}{c}{($n_p$)}           &    \multicolumn{4}{c}{($n_g$)}     \\
      ($U$)        &  MA        &  Std-GA  & GA-SJ & MA-LV       &   MA  &  Std-GA  & GA-SJ  &   MA-LV   \\
\hline
10 &  60  & 300 &  75     &  70  & 250  & 500  & 275   & 250  \\
\hline
15   &  150   & 750   &  190 &   170  &  300   &   1000  & 350 & 300  \\
\hline
20     &    400  &    2000  & 500 &  430  &  400  & 1500  & 500 & 400  \\
\hline
\end{tabular}
\label{tabla2}
\end{table}

Notice that, in general, population size of GA-SJ is $\sim 75\%$ less than that of Std-GA, and proposed MA uses $\sim 20-25\%$ less individuals than GA-SJ. MA-LV requires $\sim 10-15\%$ more individuals than proposed MA. If we look at the number of generations, both proposed MA and MA-LV reduce this parameter $\sim 10-20\%$ with respect to GA-SJ, and $\sim 50-70\%$ with respect to Std-GA.


Results in Table \ref{tabla2} were obtained adjusting parameters of the different algorithms in order to achieve a similar performance than that of the MA, so as to make fair comparisons. This is accomplished  by adjusting the number of generations, $n_g$, and population size, $n_p$.

In this case, Wilcoxon test \cite{Derrac11} between MA and Std-GA, and between MA and MA-LV. was run. This test finds significant performance  differences between two methods. In the first case, the test yields a $p$-value of 0.0398, showing again a significant improvement of MA with respect to Std-GA at the 0.1 level of significance. When the test was applied to proposed MA and MA-LV the $p$ value was 0.112, showing that MA is not significantly better than MA-LV, though MA gets a better ranking than MA-LV.

\subsection{Channel estimation accuracy}

\begin{figure}[htb]
\centerline{\psfig{file=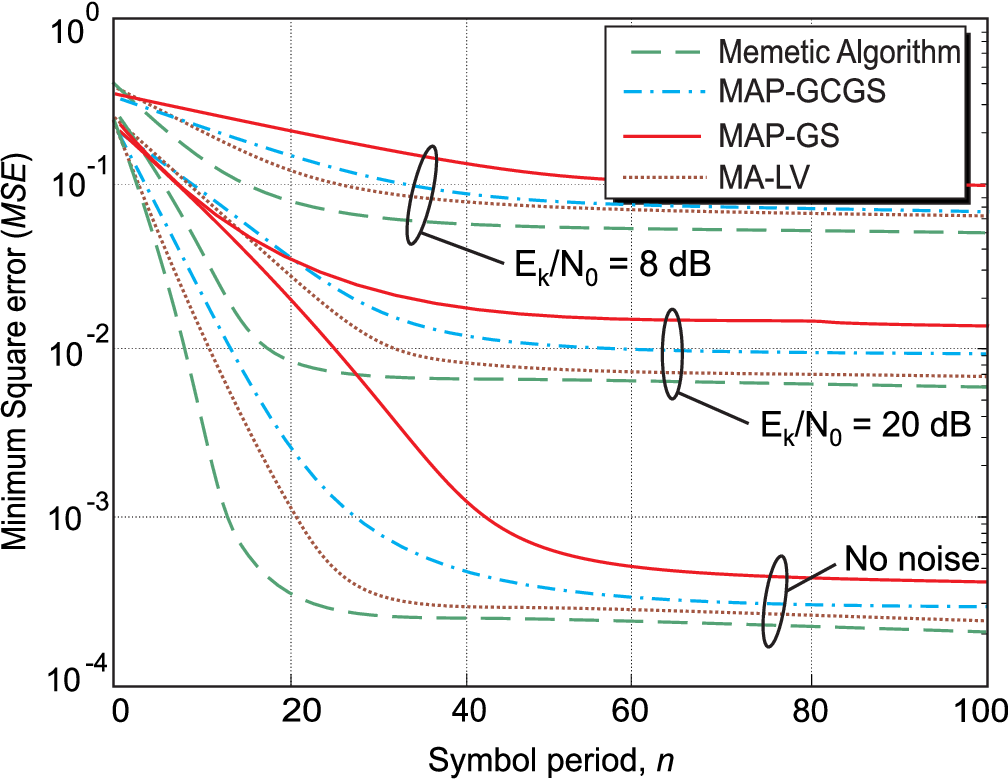,width=8cm}}
\caption{Average MSE of the channel estimate over the first 100 known symbols. Mean values obtained after 50 symbol frames. The energy received for every transmitting users is the same. $U=8$ active users.}   
\label{f5y}
\end{figure}

Next, the accuracy of the fading coefficients estimates is analyzed. Figure \ref{f5y} shows the minimum square error plots for four different schemes (proposed MA, MA-LV in \cite{Lim04}, and the ``MAP-GCGS'' and ``MAP-GS'' algorithms developed in \cite{Huang02}), for several $E_k/N_0$ values measured along a 100 symbols long frame. The number of active users is $U=8$, all of them transmitting with the same power.  Numerical results show that proposed MA requires, on average, approximately 20-30 symbol periods to achieve low error estimates, while MAP-GS and MAP-GCGS require about  45-60 samples, and the MA-LV detector requires a period of about 35-50 samples. 

An additional advantage of both MAs is that no supervised initial period is required. Conversely, both MAP-GCGS and MAP-GS need  a 18 samples long training period.

\subsection{Near-far performance}
\label{nearfar}

The near-far effect reflects the situation when part of the  interfering users are closer to the base-station than the user of interest. This involves that their signals are received with stronger power than the one coming from the user of interest, making, this way, more difficult the problem of extracting useful data \cite{Proakis98,Verdu98}. The UOI's BER as a function of the difference between the power received from the UOI and from the other users is now evaluated. Three interfering users ($U=4$), $N=32$ and noise variance $\sigma_n^2 = 0.5$, are considered. Power from the three interfering users is the same at reception antenna.

\begin{figure}[htb]
\centerline{\psfig{file=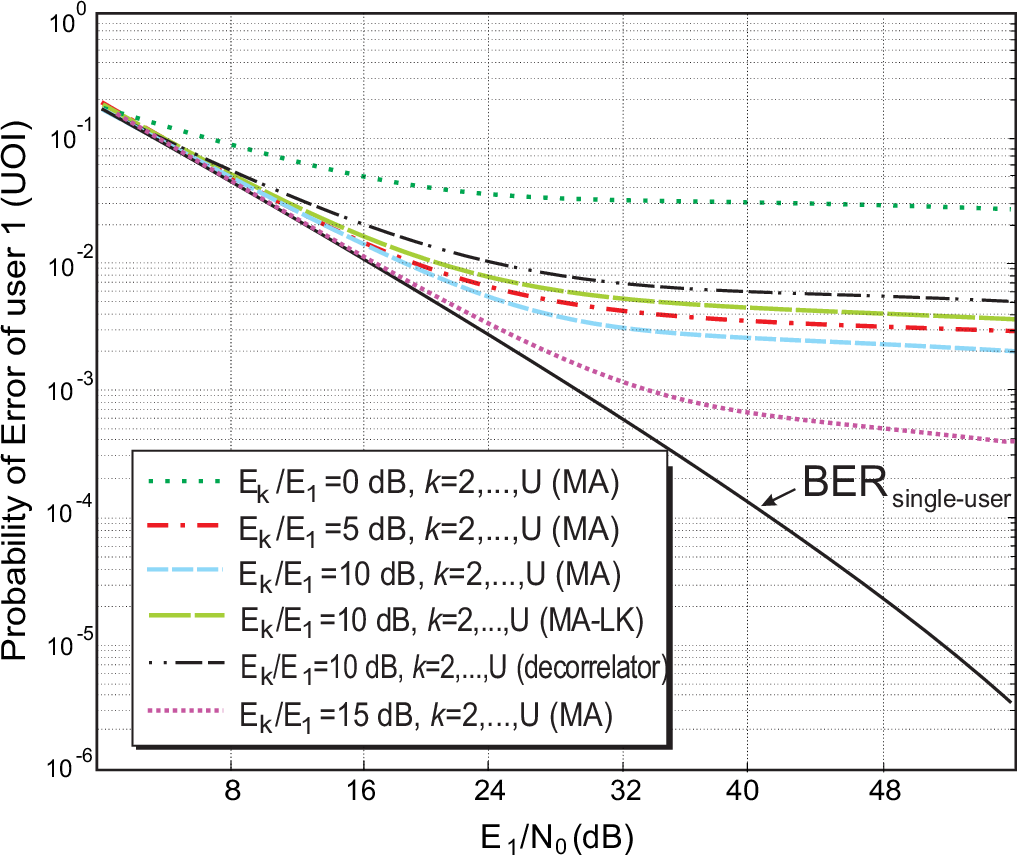,width=8cm}}
\caption{BER performance for $K=10$ users with $E_k/E_1=0$, 5, 10 and 15 dB for $k=2,\dots,U$. User 1: user of interest.}  
\label{fig_nearfar2new}
\end{figure}

Figure \ref{fig_nearfar2new} shows the near-far performance of our MA depending on the SNR of the UOI. The mean received symbol energies of the remaining users are 0, 5, 10 and 15 dB higher than the UOI's energy. Numerical results show that when $5 \leq E_k/E_1 \leq 10$ dB, MA detector has good near-far capabilities up to $E_1/N_0 \approx 18$ dB. Higher values of $E_1/E_0$ lead to a slight BER degradation.

The BER plots corresponding to: (i) the decorrelator detector, which is known to be near-far resistant, and (ii) the MA-LV algorithm, are also shown for $E_k/E_1=10$ dB to allow comparison. It can be seen  how the decorrelator performance notably decreases with respect to MA for the same $E_k/E_1$ value, while MA-LV shows an intermediate performance.

This way, we demonstrate that our proposed multiuser detector is robust and near-far resistant when both inter-symbol and multi-access interferences exist. Besides, improvements with respect to MA-LV --see end of section \ref{sec_intro}-- are proved to lead to better near-far resistance.

\section{Conclusions}
\label{sec_concl}

A memetic algorithm based multiuser detector has been developed so as to properly estimate both the channel response and the transmitted symbols in a multiuser DS/CDMA scenario, which is one of the most important and studied problems in modern communications.

Numerical experiments were carried out to compare the proposed MA MUD to different nature-inspired  detectors, as far as we know, in fair terms, including comparisons to many well-known traditional schemes such as the matched filter, the decorrelator  and even some Bayesian approaches, as well. Proposed MA is an efficient alternative when approaching this complex optimization problem, offering  a remarkable performance, especially under unfavorable conditions (e.g. low signal-to-noise power ratio of the user of interest, considerable number of interferers, or existence of near-far effects).

Whenever these degradations are not high, other two-stage nature-inspired algorithms show a performance similar to that of the MA detector and the optimum receiver. When ISI or MAI become stronger or the UOIs power is smaller, proposed MA achieves a better performance, in part as a result of  its efficient search capabilities that adjust population diversity by in-service monitoring the population fitness entropy. This way, probabilities of crossover and mutation are fine-tuned using this information, thus adapting the exploitative/explorative sense of the search.

On the other hand, the analysis of statistical significance carried out using both Friedman and Wilcoxon tests reveals that MA clearly outperforms the Genetic Algorithm approach, and also has better scores than other memetic algorithms shown for comparison. 

The obtained multiuser detector allows an efficient use of band-limited real channels. Transmission rates can be increased --even when the number of actively transmitting users is high, and/or when  fading and interference become stronger-- more than with previous approaches, at the expense of, in worst case, limited computation increments.






\vfill\pagebreak



\end{document}